\input{aipcheck}

\documentclass[
    ,final            % use final for the camera ready runs
  ]
  {aipproc}

\layoutstyle{6x9}

\usepackage{amsmath}
\usepackage{amssymb}
\usepackage{showlabels}

\newcommand{\bea}{\begin{eqnarray}}
\newcommand{\eea}{\end{eqnarray}}
\newcommand{\nn}{\nonumber}

\font\tenscr=rsfs10 scaled1100
\font\sevenscr=rsfs7 % scaled \magstep1
\font\fivescr=rsfs5 % scaled \magstep1
\skewchar\tenscr='177
\skewchar\sevenscr='177
\skewchar\fivescr='177
\newfam\scrfam
\textfont\scrfam=\tenscr
\scriptfont\scrfam=\sevenscr
\scriptscriptfont\scrfam=\fivescr

\def\scri{{\fam\scrfam I}}
\def\scre{{\fam\scrfam E}}

\begin{document}

\title{Towards a cross-correlation approach to strong-field dynamics in Black Hole spacetimes}


\author{J.L. Jaramillo}{
  address={Max-Planck-Institut f{\"u}r Gravitationsphysik, Albert Einstein
Institut, Potsdam, Germany}
}

\author{R.P. Macedo}{
  address={Theoretisch-Physikalisches Institut, Friedrich-Schiller-Universit{\"a}t Jena, Jena, Germany}
}

\author{P. Moesta}{
   address={Max-Planck-Institut f{\"u}r Gravitationsphysik, Albert Einstein
Institut, Potsdam, Germany} 
}

\author{L. Rezzolla}{
  address={Max-Planck-Institut f{\"u}r Gravitationsphysik, Albert Einstein
Institut, Potsdam, Germany}
  ,altaddress={Department of Physics and Astronomy,
Louisiana State University,
Baton~Rouge, LA, USA} % additional visiting address
}

\begin{abstract}
The qualitative and quantitative understanding of near-horizon 
gravitational dynamics in the strong-field regime represents 
a challenge both at a fundamental level and in astrophysical 
applications. Recent advances in numerical relativity and in 
the geometric characterization of black hole horizons open 
new conceptual and technical avenues into the problem. We 
discuss here a research methodology in which spacetime 
dynamics is probed through the cross-correlation of geometric 
quantities constructed on the black hole horizon and on null 
infinity. These two hypersurfaces respond to evolving
gravitational fields in the bulk, providing canonical 
"test screens" in a "scattering"-like perspective onto 
spacetime dynamics. More specifically, we adopt a 3+1 Initial 
Value Problem approach to the construction of generic spacetimes and 
discuss the role and properties of dynamical trapping 
horizons as canonical inner "screens" in this context. We 
apply these ideas and techniques to the study of the recoil 
dynamics in post-merger binary black holes, an important issue 
in supermassive galactic black hole mergers.

\end{abstract}

\maketitle

%%%%%%%%%%%%%%%%%%%%%%%%%%%%%%%%%%%%%%%%%%%%
%% MAINMATTER
%%%%%%%%%%%%%%%%%%%%%%%%%%%%%%%%%%%%%%%%%%%%

\section{A cross-correlation approach: Motivations and Objective}
The general problem here discussed is the qualitative and 
quantitative understanding of near-horizon gravitational dynamics in the 
strong-field regime of black hole (BH) spacetimes. 
This represents a  challenge both at a 
fundamental level and in astrophysical applications.
The setting of our discussion is that of classical spacetimes in 
General Relativity (GR), with a  focus on astrophysically motivated 
problems in which we adopt a numerical relativity methodology.

A natural strategy to the study of spacetime dynamics
consists in extending to the
general relativistic setting the Newtonian description of 
the dynamics of gravitationally interacting bodies, namely
celestial mechanics.
This has proved extremely successful in unveiling the physics
of compact objects. However, such an approach also
meets fundamental obstacles in the general dynamical regime
of a gravitational theory in which i) {\em a priori} rigid structures
providing canonical references (such as symmetries or preferred backgrounds) 
are generically absent, and ii) where 
global aspects play a crucial role.
Here we rather adopt a complementary approach in the spirit of a 
{\em coarse-grained}
description of the dynamics in which  we renounce  to the detailed tracking 
of the geometry (trajectories) of given compact regions
(objects), and rather
emphasize  the {\em global/quasi-local properties} of the relevant 
dynamical fields.
In particular, we aim at capturing the functional structure of the latter
through 
appropriate {\em correlation functions}.

From a physical perspective, we focus on what is sometimes called the 
``establishment's picture of gravitational collapse'' \cite{Pen73}. 
Our current  understanding of this problem can be summarized in the following
chain of theorems and conjectures:
\begin{itemize} 
\item[i)] Singularity 
theorems \cite{Pen65b,Hawki67,HawPen70,HawEll73}: 
if sufficient {\em energy} is placed in a sufficiently 
compact spacetime region, light locally  converges in any emitted
direction, trapped surfaces form and a spacetime singularity develops in their
causal future.
\item[ii)] Weak cosmic censorship conjecture \cite{Pen69}:
to preserve predictability, the singularity is hidden from
a distant observer behind
an event horizon giving rise to a BH region.
\item[iii)] BH spacetime
stability (conjecture): GR dynamics drive the system to stationarity.
\item[iv)] BH uniqueness (theorem) \cite{Heusl98}:
the final state is a subextremal Kerr BH spacetime.
\end{itemize}
The establishment's picture of gravitational collapse is an intrinsically
dynamical picture. Accordingly, we adopt a methodology able
to cope with the construction and analysis of generic spacetimes, namely
in an Initial Value Problem approach. This provides a 
systematic avenue to the study of the  qualitative and quantitative aspects
of generic dynamical spacetimes, respectively addressed through the use
of tools in Partial Differential Equation theory and by
the numerical construction of spacetimes.

Our focus is on the point  iii) above, in an attempt to gain insight into
the specific manner in which GR drives the system  eventually to stationarity.
Remarkably, tools devised in the study of i) and ii) 
(namely, the characterizations of BH horizons) prove to be very useful 
to reach this goal.
In this sense, a crucial outcome of the systematic 
numerical exploration of fully dynamical vacuum spacetimes in recent years
is that the a posteriori description of {\em  gravitational dynamics is rather simple}.
This specific observation is the main point supporting the applicability
of a coarse-grained approach to the analysis
of generic spacetime dynamics.

\subsection{A cross-correlation approach to BH spacetime dynamics}
\begin{figure}
   \includegraphics[height=.3\textheight]{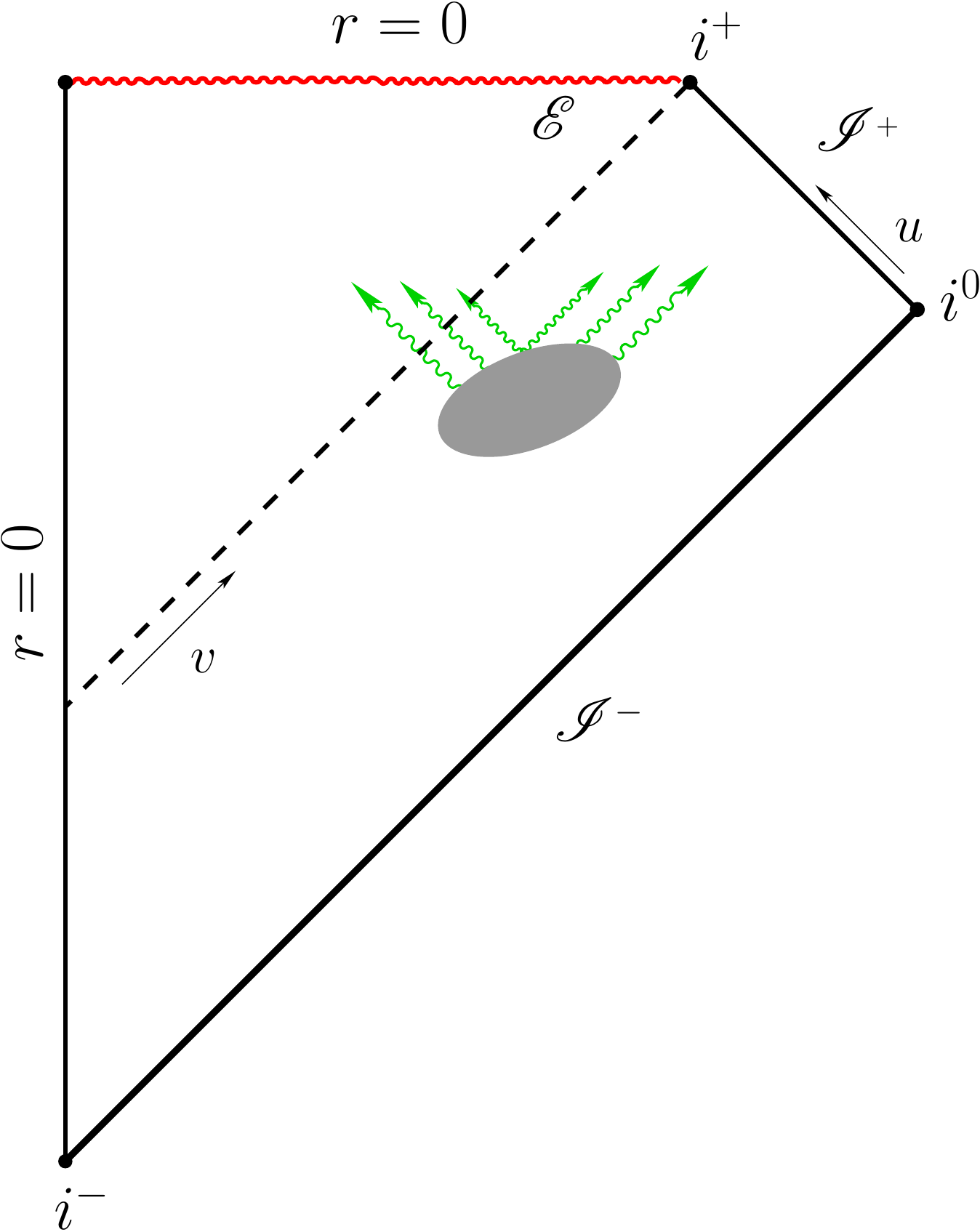}
   \label{f:BHscattering}
   \caption{Sketch of the conformal diagram for a spherically symmetric
            gravitational collapse, illustrating the scenario
            for the cross-correlation approach to the analysis of 
            bulk spacetime dynamics.}
 \end{figure}

In order to make concrete the previous considerations, we formulate the 
following {\em cross-correlation} methodology 
\cite{Jaramillo:2011re,Jaramillo:2011rf,Jaramillo:2011zw}, whose specific goal
is the development of qualitative insights into spacetime dynamics
in the strong-field regime, identifying the key elements 
leading ultimately to appropriate quantitative 
effective descriptions:
\begin{itemize}
\item[i)] Spacetime dynamics is 
probed through the cross-correlation of geometric quantities
$h_{\mathrm{inn}}$ an $h_{\mathrm{out}}$ defined
at {\em inner} and {\em outer}  hypersurfaces, respectively, 
${\cal H}_{\mathrm{inn}}$ and ${\cal H}_{\mathrm{out}}$. 
\item[ii)] Hypersurfaces ${\cal H}_{\mathrm{inn}}$ and ${\cal H}_{\mathrm{out}}$
are taken as {\em test screens} responding to bulk dynamics.
Spacetime is then explored in the spirit of an {\em "inverse scattering approach"}.
\end{itemize}
In our near-horizon and asymptotically flat context, the  BH event horizon 
${\cal E}$ and future null infinity $\scri^+$ provide natural choices, respectively,
for ${\cal H}_{\mathrm{inn}}$ and ${\cal H}_{\mathrm{out}}$
(cf. Fig. \ref{f:BHscattering}). 
Quantities $h_{\mathrm{inn}}(v)$ 
and $h_{\mathrm{out}}(u)$ are then constructed on ${\cal E}$ and  $\scri^+$ 
as functions, respectively, of appropriate advanced and retarded times.
We note that the cross-correlations (as time-series)
between $h_{\mathrm{inn}}(v)$ and $h_{\mathrm{out}}(u)$ 
require a {\em gauge mapping} between $v$ and $u$.

%% \begin{figure}[t!]
%% \begin{center}
%% \vglue-1.0cm
%% \includegraphics[angle=90,width=13.0cm,clip=true,angle=-90]{fig2.eps}
%% \end{center}
%% \vglue-3.75cm
%% \caption{Carter-Penrose diagram illustrating the {\em scattering}
%%   approach to near-horizon gravitational dynamics in a generic
%%   spherically symmetric collapse. The event horizon $\scre$ and
%%   null infinity $\scri^+$ provide spacetime canonical screens on which
%%   {\em geometric quantities}, respectively accounting for horizon
%%   deformations and wave emission, are defined. Their cross-correlation
%%   encodes nontrivially information about the bulk spacetime dynamics.}
%% \label{fig:BHscattering}
%% \end{figure}

\section{BH horizons: Global vs. Quasi-local approaches}

BH horizons play a crucial role in our discussion, 
offering in particular a model for the inner hypersurface ${\cal H}_{\mathrm{inn}}$  
in the cross-correlation scheme. Two approaches to the 
BH notion can be considered, both contained in the the standard picture 
of gravitational collapse.
The first one is guaranteed by weak cosmic censorship: the black hole region 
$\mathcal{B}$ is a region of {\em no-escape} not extending up to infinity. 
In a strongly asymptotically predictable spacetime ${\cal M}$, with
 $J^-(\scri^+)$ denoting the causal past of
$\scri^+$, we have  
$\mathcal{B} = {\cal M} - J^-(\scri^+)$. The boundary of the 
black region is given by the event horizon 
$\mathcal{E} = {\partial{J}}^-(\scri^+)\cap {\cal M}$.
This traditional characterization of BHs for asymptotically
flat spacetimes involves global spacetime concepts. In particular,
event horizons $\mathcal{E}$ are teleological objects whose location requires
the full knowledge of the spacetime in the future and that
can develop in flat regions. Event horizons are therefore
not adapted for probing the BH spacetime during its construction
in an Initial Value Problem approach.

Alternatively, instead of characterizing the BH as the region that cannot
send signals to distant observers, we can approach it as the region where
all emitted light rays ``locally converge''. This is made precise by the notion
of trapped surface, crucial in the singularity theorems appearing in the 
standard gravitational collapse picture. Given a closed  
surface ${\cal S}\subset {\cal M}$
with area element $dA =\sqrt{q}\;d^2x$, its
normal plane can be spanned in terms of an outgoing null vector 
$\ell^a$ and an ingoing null vector $k^a$. The outgoing expansion $\theta^{(\ell)}$ 
measures the rate of change of $dA$ in the lightfronts emitted
from ${\cal S}$ along the outgoing null direction $\ell^a$. The ingoing expansion
$\theta^{(k)}$ is defined analogously
\bea
\label{e:expansion}
\theta^{(\ell)}\equiv \frac{1}{\sqrt{q}}{\cal L}_{\ell} \sqrt{q}
\ \ \ , \ \ \
\theta^{(k)}\equiv \frac{1}{\sqrt{q}}{\cal L}_{k} 
\sqrt{q} \ \ .
\eea
Then ${\cal S}$ is a (future) trapped surface \cite{Pen65b} if $\theta^{(\ell)}<0$
and $\theta^{(k)}<0$. The limiting case in which one of the
expansion vanishes defines a marginally trapped surface:  
$\theta^{(\ell)}=0$, $\theta^{(k)}<0$. If there exists a notion of {\em outer region},
e.g. associated with an asymptotically flat region, then  
(future) outer trapped surfaces \cite{HawEll73} can be introduced
as $\theta^{(\ell)}<0$, without any requirement on $\theta^{(k)}$.
A marginally outer trapped surface satisfies then $\theta^{(\ell)}=0$.
In this context the BH can be characterized in terms of the {\em trapped region}
${\cal T}$,
namely the set of points in spacetime belonging to some trapped surface.
In the present context of a cross-correlation analysis,
rather than in the BH or trapped region themselves
we are primarily interested in its boundary. 
 The latter provides a hypersurface to be employed as an inner screen 
${\cal H}_{\mathrm{inn}}$. In this sense,
the so-called {\em trapping boundary} \cite{Hay94}, namely the boundary of 
the trapped region ${\cal T}$,  would be a good candidate for ${\cal H}_{\mathrm{inn}}$. 
Unfortunately, we lack an operational 
characterization of such trapping boundary. This is in contrast
with the notion of {\em apparent horizon}, namely the boundary of the trapped
region contained in a given 3-slice $\Sigma$, characterized
as a marginally outer trapped surface $\theta^{(\ell)}=0$. 
Motivated by these difficulties, trapping horizons (namely
worldtubes of marginally outer trapped surfaces) were introduced
\cite{Hay94} as quasi-local models for the BH horizon, in particular
in an attempt to gain insight about the trapping boundary.

\section{Quasi-local horizons as inner test screens}
We review the properties of dynamical trapping horizons, stressing
those aspects of special relevance in our discussion.
Let us consider a closed orientable 2-surface
${\cal S}$ embedded in a 4-dimensional spacetime 
$({\cal M}, g_{ab})$, with Levi-Civita connection $\nabla_a$.
Regarding its intrinsic geometry, we denote the induced metric 
as $q_{ab}$, with Levi-Civita connection $D_a$, Ricci scalar
${}^2\!R$ and area form  $\epsilon_{ab}$ (area measure $dA$).
Again, normal outgoing and ingoing null normals are 
denoted as $\ell^a$ and $k^a$, normalized as $\ell^a k_a = -1$.
This leaves  a (boost) rescaling freedom
  $\ell'^a =f \ell^a$, $k'^a = f^{-1} k^a$, with $f$ a function on ${\cal S}$.

We need the following elements of the extrinsic geometry.
The outgoing expansion $\theta^{(\ell)}$, given in Eq. (\ref{e:expansion}), and 
the outgoing shear $\sigma^{(\ell)}_{ab}$ are expressed as   
\bea
\label{e:expansion_shear}
\theta^{(\ell)}=q^{ab}\nabla_a\ell_b \ \ \ \ , \ \ \ \ 
\sigma^{(\ell)}_{ab}=  {q^c}_a {q^d}_b \nabla_c \ell_d - \frac{1}{2}\theta^{(\ell)}q_{ab} \ , 
\eea
whereas a normal fundamental 1-form $\Omega^{(\ell)}_a$
associated with $\ell^a$ 
is given by
\bea
\label{e:Omega}
\Omega^{(\ell)}_a = -k^c {q^d}_a \nabla_d \ell_c \ ,
\eea
provides a connection on $T^\perp{\cal S}$.
The transformation  of these quantities under a null rescaling are:
$\theta^{(\ell')}=f \theta^{(\ell)}$, $\sigma^{(\ell')}_{ab}= f\sigma^{(\ell)}_{ab}$
and $\Omega^{(\ell')}_a = \Omega^{(\ell)}_a + D_a(\mathrm{ln}f)$.
Finally, we need to control the variations of the outgoing expansion $\theta^{(\ell)}$
along normal vectors $v^a\in T^\perp{\cal S}$
\bea
\label{e:delta_theta}
\delta_{\alpha \ell} \theta^{(\ell)} &=& \kappa^{(\alpha \ell)} \theta^{(\ell)}
- \alpha\left[\sigma^{(\ell)}_{ab} {\sigma^{(\ell)}}^{ab} + 
G_{ab} \ell^ak^b + \frac{1}{2}  \left(\theta^{(\ell)}\right)^2 \right] \nn
\\
\delta_{\beta k} \theta^{(\ell)} &=& \kappa^{(\beta k)} \theta^{(\ell)}
+ {}^2\!\Delta \beta - 2 \Omega^{(\ell)}_a  D^a\beta \nn \\
&&+\beta \left[ \Omega^{(\ell)}_a  {\Omega^{(\ell)}}^a 
- D^a  \Omega^{(\ell)}_a  -\frac{1}{2}{}^2\!R + G_{ab}k^a\ell^b 
-\theta^{(\ell)}\theta^{(k)} \right]  \ ,
\eea
where $\kappa^{(v)} = - v^a k^b \nabla_a \ell_b$, $\alpha$ and $\beta$ are functions on ${\cal S}$ and 
$\delta_v$ is the variation operator associated with a change
in the underlying surface ${\cal S}$ (cf. \cite{AndMarSim05,AndMarSim07,BooFai07}).

A {\em trapping horizon} \cite{Hay94} is (the closure of) a hypersurface ${\cal H}$ 
foliated by closed marginally outer trapped surfaces: 
${\cal H} = \bigcup_{t\in\mathbb{R}} {\cal S}_t$,  
with  $\theta^{(\ell)}=0$. The properties of ${\cal H}$ as a horizon
are characterized by the signs of $\theta^{(k)}$ and $\delta_{k}\theta^{(\ell)}$:
i) the sign of   $\theta^{(k)}$ controls if the singularity occurs in 
 the  {\em future} ($\theta^{(k)}<0)$ or in the {\em past} ($\theta^{(k)}>0)$, and ii)
the sign of $\delta_{k}\theta^{(\ell)}$ controls 
the (local) {\em outer} ($\delta_{k}\theta^{(\ell)}<0$) 
or {\em inner} ($\delta_{k}\theta^{(\ell)}>0$) character of ${\cal H}$ with respect
to the trapped region. For BHs the singularity occurs in the future
and the horizon is an outer boundary. Therefore quasi-local BH horizons
are modeled by {\em future outer trapping horizons} (FOTH): $\theta^{(\ell)}=0$,
$\theta^{(k)}<0$ and $\delta_{k}\theta^{(\ell)}<0$.

We can define an evolution vector $h^a$ on ${\cal H}$
characterized by: i)  $h^a$ is tangent to ${\cal H}$ and orthogonal
to ${\cal S}_t$, ii) $h^a$ transports ${\cal S}_t$ to ${\cal S}_{t+\delta t}$:
${\cal L}_h t =1$, and iii)   $h^a$ is written as $h^a = \ell^a - C k^a$. 
We also define a {\em dual} vector  $\tau^a= \ell^a + C k^a$ orthogonal 
to ${\cal H}$.
The sign of $C$ fixes the point-like  metric type of ${\cal H}$: 
$C>0$ spacelike, $C=0$ null, $C<0$ timelike.
Considering first the spherically symmetric case 
($C$ constant on ${\cal S}_t$), the trapping horizon
conditions  $\theta^{(\ell)} = 0$, $\delta_h \theta^{(\ell)} = 0$ imply, using
Eq. (\ref{e:delta_theta}) and Einstein equations
\bea
\label{e:C_spher}
C = - \frac{\sigma_{ab}^{(\ell)}{\sigma^{(\ell)}}^{ab}+ T_{ab}\ell^a\ell^b}
{\delta_k \theta^{(\ell)}} \geq 0 \ ,
\eea 
for an outer ${\cal H}$ (i.e. $\delta_k \theta^{(\ell)}<0$)
under the null energy condition. Therefore FOTHs ${\cal H}$ can be either
null or spacelike hypersurfaces. The first case corresponds to the 
stationary regime (with its the {\em isolated horizon} hierarchy 
\cite{AshKri04,GouJar06}),
whereas the second case leads to {\em dynamical horizons} (DHs) 
\cite{AshKri02,AshKri03}.
Beyond spherical symmetry, one can pose the question 
if a MOTS section ${\cal S}_t$ of a 
FOTH can be partially spacelike and partially null, i.e. if it can happen 
$C>0$ in a part of ${\cal S}_t$  and $C=0$ in another part.
The answer is in the negative: transitions from stationarity to the dynamical
regime happens ``all at once'' in sections ${\cal S}_t$ of ${\cal H}$. 
This follows from the trapping horizon condition written as [use (\ref{e:delta_theta})] 
\bea
\label{e:stability_operator}
 \delta_h \theta^{(\ell)} = -{}^2\!D_c {}^2\!D^c C
 + 2 {\Omega^{(\ell)}}^c {}^2\!D_c C -C \delta_k\theta^{(\ell)} + 
 \delta_\ell\theta^{(\ell)} = 0 \ .
\eea
Under an outer condition $\delta_k\theta^{(\ell)}<0$, a maximum
principle can be applied to this elliptic equation to conclude
that either $C>0$ (if 
$\delta_\ell\theta^{(\ell)}=\sigma_{ab}^{(\ell)}{\sigma^{(\ell)}}^{ab}+ T_{ab}\ell^a\ell^b
\neq 0$ at some point of ${\cal S}_t$), or  $C=0$ if and only 
if $\delta_\ell\theta^{(\ell)}=0$ everywhere on
${\cal S}_t$. In other words, it suffices that some energy crosses
the horizon at a single point, for making the whole horizon grow
globally. This non-local behaviour is encoded in the elliptic 
nature of Eq. (\ref{e:stability_operator}), providing an example
of the non-local behaviour of these dynamical trapping horizons.
This is possibly disturbing, 
if considering ${\cal H}$ as the boundary of a physical object.  
An even more curious behaviour of ${\cal H}$ follows from the two following results:
\begin{itemize}
\item [i)] {\em Foliation uniqueness} \cite{AshGal05}: the foliation by MOTSs
of a dynamical FOTH is  unique.
\item [ii)] {\em Existence of DHs} \cite{AndMarSim05,AndMarSim07}: 
Given 
a (stable) marginally outer trapped surface ${\cal S}_0$ in a Cauchy hypersurface
$\Sigma_0$, to each 3+1 spacetime foliation $(\Sigma_t)_{t\in\mathbb{R}}$ 
containing $\Sigma_0$ it
corresponds a unique adapted dynamical FOTHs ${\cal H}$
that contains  ${\cal S}_0$ and is sliced by marginally outer trapped surfaces
$\{ {\cal S}_t \}$ such that
${\cal S}_t\subset\Sigma_t$.
\end{itemize}
The first result provides a sort of {\em rigidity} for DHs, 
very useful in our context since it determines the 
evolution vector $h^a$ uniquely up to a time reparametrization.
Regarding the second result, this is a crucial benchmark in the 
treatment of quasi-local horizons in an Initial Value Problem approach.
The main point we want to underline here is that the combination of
results i) and ii) above leads to the non-uniqueness of DHs: 
the combination of results on {\em evolution existence}
and {\em foliation uniqueness} for DHs
implies the generic {\em non-uniqueness}
in the evolution of a FOTH from
an initial MOTS. To see this it suffices to consider the evolution of an
initial ${\cal S}_0\in \Sigma_0$ into DHs ${\cal H}_1$ and ${\cal H}_2$
compatible with (generic) 3+1 foliations $\{\Sigma_{t_1}\}$ and $\{\Sigma_{t_2}\}$,
assume  ${\cal H}_1={\cal H}_2$ and then reason by contradiction 
with the result on the uniqueness of the foliation (see e.g. \cite{Jaramillo:2011rf}).
Such non-uniqueness in the evolution is a surprising behaviour
if we consider ${\cal H}$ as the boundary of a physical
object. Actually such behaviour is a characteristic
signature of gauge dynamics. At this point it is worth to remark that the
amount of gauge freedom in the stationary and the dynamical cases remains the same:
whereas i) in equilibrium the geometric hyperfurface ${\cal H}$ is unique,
but its foliation by MOTS is not unique and can be parametrized by 
a free function $f$ on ${\cal S}$ rescaling the null normal 
($\ell^a \rightarrow f \ell^a$), ii) in the dynamical case the geometric
hypersurface is non-unique but the foliation by MOTS is rigid. In this latter case,
the gauge freedom is given by the choice 3+1 foliation, i.e.
by the lapse function $N$ evaluated on ${\cal S}$. Therefore, both 
equilibrium and dynamical cases present the gauge freedom (the choice
of a function on ${\cal S}$), but dressed differently.

\subsection{A pragmatical view on FOTHs}

From the discussion above we must retain: i) FOTHs are hypersurfaces
tracking the BH region, ii) they are well-adapted to the 3+1 Initial Value Problem
approach, and iii) DHs
incorporate a sort of rigidity fixing the evolution vector up to
time reparametrization.
These are remarkable geometric properties.
If considered as physical surfaces, FOTHs present curious
properties: i) non-uniqueness, ii) superluminal behaviour (when dynamical) and
iii) global behaviour. The last point has been
addressed in detail in Ref. \cite{Bengtsson:2010tj}, where {\em clairvoyance}
properties of trapped surfaces are discussed, in particular their capacity to
enter into flat regions, one of the reasons 
to abandon event horizons.

Although dynamical trapping horizons have proved very useful to gain crucial insight
into physical aspects such as BH thermodynamics in 
dynamical contexts \cite{AshKri04,Hay94}, for the reasons 
discussed above we adopt 
here a perspective underlining the role of FOTHs as purely geometric
{\em probes} to be employed the analysis of dynamical spacetimes.
When considering an inner test screen
${\cal H}_{\mathrm{inn}}$ in the context of the cross-correlation approach, we
look for a hypersurface that: i) should be a footprint of the BH presence, 
in particular providing a probe into their spacetime
geometry, ii) should be suitable for the Initial Value Problem approach, 
providing in this setting a preferred
geometrically defined structure with some sort of {\em rigidity}, and iii) 
should intrinsically incorporate an evolution concept, tracking
the evolution of the BH properties.
In this context, {\em we adopt a pragmatical approach on DHs, as hypersurfaces 
of remarkable geometric  properties 
in BH spacetimes, providing preferred geometric probes into 
the BH spacetime geometry}.

\subsection{Initial Value Problem and cross-correlation approaches}

The adoption of an Initial Value Problem approach to the 
construction of the spacetime has an impact on the cross-correlation
scheme sketched above. On the one hand, as already discussed,
the event horizon $\scre$ is not available during the evolution. 
In addition, in a Cauchy evolution of an asymptotically flat spacetime
one does not constructs null infinity $\scri^+$, but rather the
slices $\Sigma_t$ reach spatial infinity $i^0$ (as an alternative, one
could solve a hyperboloidal Initial Boundary Value Problem, rather
than a Cauchy one, to construct $\scri^+$).
On the other hand, advanced and retarded times $u$ and $v$ are not 
natural, and one rather employ a 3+1 time function $t$ associated
with the slicing $\{\Sigma_t\}$.

\begin{figure}
   \includegraphics[height=.3\textheight]{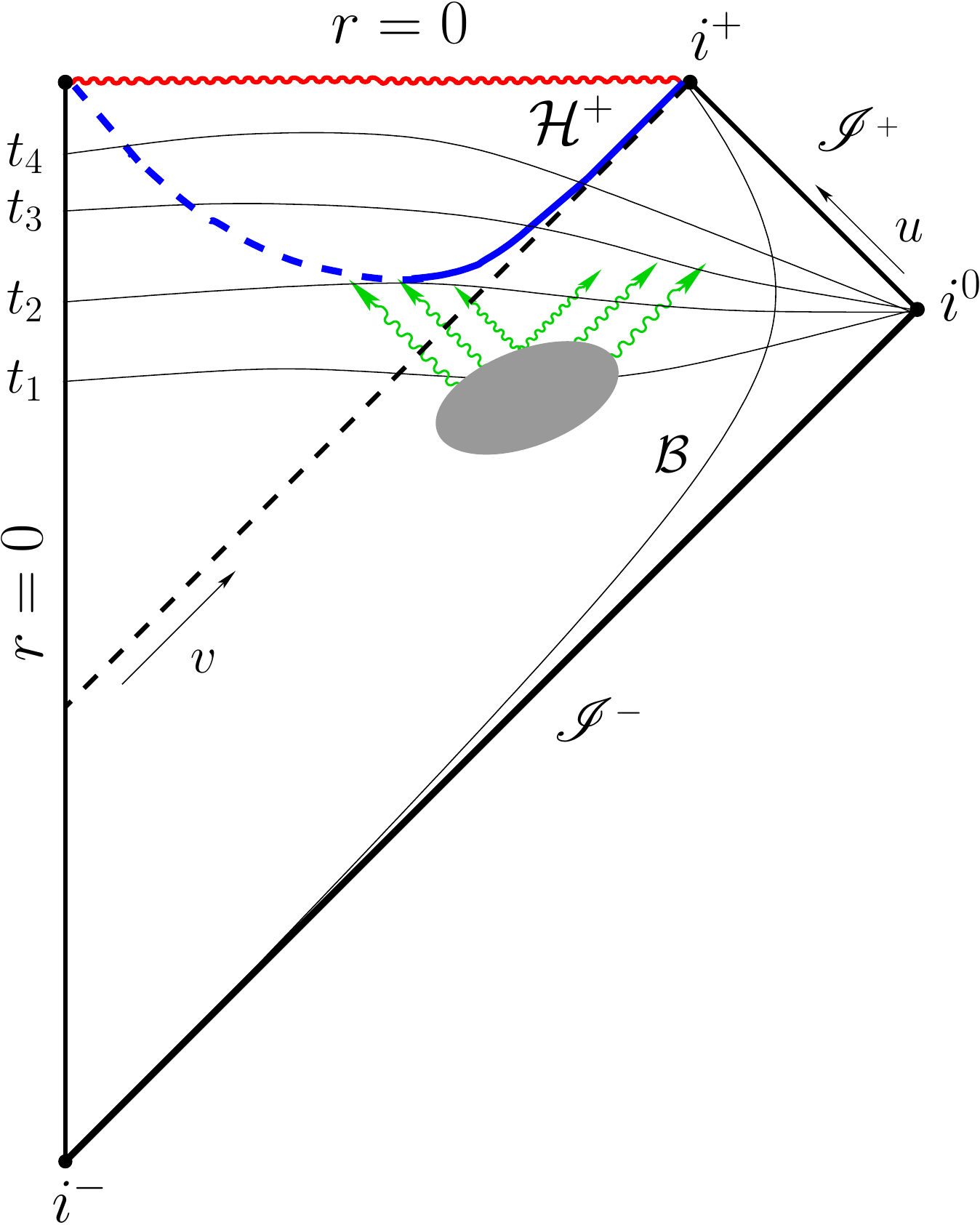}
   \label{f:3+1BHscattering}
   \caption{Illustration of the use of the Initial Value Problem approach 
    in the cross-correlation scheme. A DH is used as inner screen 
    ${\cal H}_{\mathrm {inn}}$, whereas the large spheres 
    world-tube ${\cal B}$ (or  $\scri^+$) is used for ${\cal H}_{\mathrm {out}}$.}
 \end{figure}

Therefore, in our 3+1 treatment of the cross-correlation 
methodology we shall employ (cf. Fig. \ref{f:3+1BHscattering}): 
i) dynamical trapping horizons as
canonical inner screens
associated with the 3+1 slicing with lapse function $N$,
ii) a timelike worldtube ${\cal B}$
at large distances as outer screen (or $\scri^+$ in
a hyperboloidal slicing), and iii)
a 3+1 spacetime slicing time function $t$ that
automatically implements a (gauge) mapping
between $u$ and $v$. 

\begin{figure}
   \includegraphics[height=.3\textheight]{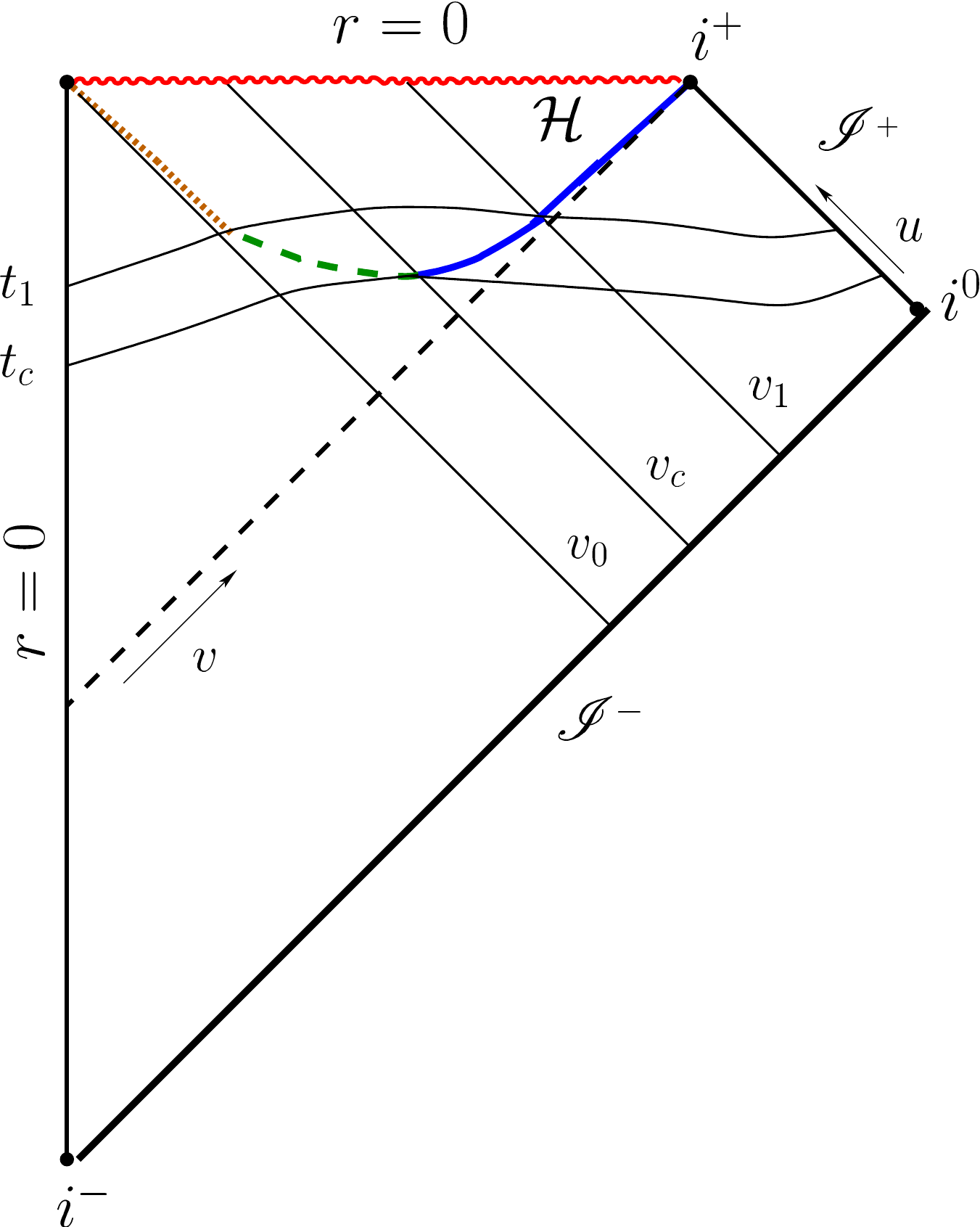}
   \label{f:InnerOuterHor}
   \caption{Internal and external apparent horizons in a 3+1 BH
    evolution. After the moment $t_c$ of first appearance,  slices $\Sigma_t$ 
    intersect multiply the DH ${\cal H}$, splitting it into ${\cal H}_{\mathrm{ext}}$  
    and ${\cal H}_{\mathrm{int}}$ (dashed line) parts.}
 \end{figure}

In the context of the cross-correlation of quantities $h_{\mathrm{inn}}(t)$
and  $h_{\mathrm{out}}(t)$ as time series, we note that
generically the 3+1 slices $\Sigma_t$ intersect 
multiply the dynamical trapping horizon (cf. Fig. \ref{f:3+1BHscattering}).
This is the underlying reason for the {\em jumps} occurring
in the evolution of apparent horizons, generic
in 3+1 BH evolutions. From the 3+1 perspective both an {\em external} 
${\cal H}_{\mathrm{ext}}$ and an {\em internal} ${\cal H}_{\mathrm{int}}$
apparent horizon are present. In numerical simulations it is 
standard to neglect the internal horizon as irrelevant.
The splitting of the single spacetime hypersurface ${\cal H}$
in two parts, ${\cal H}={\cal H}_{\mathrm{int}} \cup {\cal
  H}_{\mathrm{ext}} $  is an artifact of the 3+1
description and simply reflects that the coordinate
$t$ is not a good label for ${\cal H}$. If we are interested
in cross-correlations  only after the moment
$t_c$ of first appearance of the apparent horizon,
then  using the external part ${\cal H}_{\mathrm{ext}}$ is enough.
But if integrating fluxes in time 
is relevant in our problem, then the internal ${\cal H}_{\mathrm{int}}$
must be kept into the picture in order to account for the
whole history of the flow into the BH singularity (cf.
Fig. \ref{f:InnerOuterHor}).

%% In Fig.~\ref{fig:advanced_time} we illustrate this picture in a
%% simplified (spherically symmetric) collapse scenario that retains the
%% relevant features of the discussion. On one side, the relevant outer
%% screen boundary (namely, null infinity $\scri^+$) is parametrized by
%% the retarded time $u$, something explicitly employed in the expression
%% of the flux of Bondi momentum in Eqs. (33) and (34) of paper I. On the
%% other side, from the $3+1$ perspective, the moment $t_c$ of first
%% appearance of the (common) horizon corresponds to the coordinate time
%% $t$ at which the $3+1$ foliation $\{\Sigma_t\}$ firstly intersects the
%% dynamical horizon ${\cal H}$. For $t>t_c$, $\Sigma_t$ slices intersect
%% twice (multiply, in the generic case) the hypersurface ${\cal H}$
%% giving rise to the external and internal common horizons (\cf ${\cal
%%   H}$ in Fig.~\ref{fig:advanced_time}). Therefore, the time function
%% $t$ is not a good parameter for the whole dynamical horizon ${\cal
%%   H}$. An appropriate parametrization of this hypersurface ${\cal H}$
%% is given in terms of an advanced time, such as $v$, parametrizing past
%% null infinity $\scri^-$. 

\section{Application to Black Hole recoil dynamics}

We apply now the previous ideas to the study of the recoil
dynamics of the  BH resulting from the asymmetric
merger of two BHs. In an asymmetric binary
merger the emission of linear momentum through 
gravitational waves is not isotropic, so that the final
remnant must recoil in order to preserve the total linear
momentum.  This is astrophysically relevant
in the context of the merger of supermassive BHs
in galaxy encounters.

We aim at gaining insight into the dynamics controlling 
the recoil (kick) velocity $v_{\mathrm{k}}$ and, in particular, 
into the systematics of a late-time deceleration referred to as 
the {\em anti-kick} in the literature.
On the one hand, the very presence of such final 
deceleration is a direct consequence of the fact that
the kick velocity is obtained upon integration of
a decaying oscillating quantity, namely the flux of linear
Bondi momentum at null infinity: there is no reason
to expect that the maximum and the asymptotic value
of the time integral of a decaying oscillating signal coincide
\cite{Sperhake:2010uv}. The relevance of this problem lays, 
rather than on the presence itself of the {\em anti-kick},
on the capability to estimate {\em a priori} its
magnitude from an understanding of the underlying dynamics.
In particular, a relative large antikick reduces the resulting
kick velocity, whereas large final recoil velocities are
associated with small relative antikicks.
In this context, the magnitude of the antikick is controlled by the 
ratio between the oscillation timescale, $T$, and the 
decaying timescale, $\tau$, leading to the crucial 
notion of {\em slowness parameter}  \cite{Price:2011fm}
\bea
\label{e:slowness_parameter}
P=\frac{T}{\tau} \ \ ,
\eea
so that for $P\ll 1$ the rapid oscillations in the signal induce cancellations
in the time integral and lead to large antikicks, whereas
for $P\approx1$ the antikick is small.

Understanding the involved oscillating and decaying dynamics is therefore
crucial in this problem.
Ref. \cite{Rezzolla:2010df} paved the way to get insight into the
responsible gravitational dynamics, in terms of the analysis of the evolution
(dissipation) of the quasi-local horizon geometry. 
The cross-correlation scheme \cite{Jaramillo:2011re,Jaramillo:2011rf,Jaramillo:2011zw}
permits to develop a more systematic analysis along those lines.
Before proceeding further, and since it will be relevant later,
we note that
quasi-local BH horizon tools have already been applied 
to this problem. In particular, Ref. \cite{KriLouZlo07} proposes an expression 
for the BH linear momentum
\bea
\label{e:P_ADM}
P[\xi] = \frac{1}{8\pi}\int_{{\cal S}_t} \left(K_{ab} -K \gamma_{ab}\right)
\xi^a s^b \; dA \ ,
\eea
by applying the ADM prescription at spatial infinity $i^0$ to the apparent horizon.

\subsection{Cross-correlations in BH recoil dynamics}

In the application of the cross-correlation approach to BH recoil
dynamics, we take as quantity $h_{\mathrm{out}}$ the flux of Bondi
linear momentum at $\scri^+$  (actually at the approximation given by
large spheres  worldtube ${\cal B}$) along a spatial direction
$\xi^a$. That is
\bea
\label{e:flux_Bondi_momentum}
\frac{dP^{\mathrm{B}}[\xi]}{dt}(t) =  \lim \limits_{r \to \infty}  
\frac{r^2}{16 \pi} \oint_{{\cal S}_{t,r}} (\xi^i s_i) 
\left|{\cal N}(t)\right|^2 d\Omega
\ \ \ \ , \ \ \ \ {\cal N}(t) = 
\int_{-\infty}^t \Psi_4(t') dt'  \ ,
\eea
where ${\cal N}(t)$ is the so-called {\em news} function.
At the inner DH  screen ${\cal H}$,
we lack a geometric analogue to $dP^{\mathrm{B}}[\xi]/dt$,
in particular no news function formalism is available. However,
Eq. (\ref{e:flux_Bondi_momentum}) suggests 
a natural heuristic candidate $\tilde{K}[\xi]$ for the quantity $h_{\mathrm{inn}}$ 
\bea
\tilde{K}[\xi](t) \equiv  - \frac{1}{16 \pi}
\oint_{{\cal S}_t} (\xi^is_i) \left|\tilde{\cal N}_{\Psi}^{(\ell)}(t)\right|^2 dA \ \ \ \ , 
\ \ \ \ 
\tilde{\cal N}_{\Psi}(t) 
\equiv \int_{t_0}^v \Psi^N_0(t') dt' \ ,
\eea
where the Weyl scalar $\Psi^N_0$ plays the analogous role at ${\cal H}_{\mathrm{inn}}$
that $\Psi_4$ plays at ${\cal H}_{\mathrm{out}}$ (the superindex
$N$ refers to a choice of null tetrad adapted to the 3+1 foliation
with lapse function $N$ \cite{Jaramillo:2011rf}).
However, 
the function $\tilde{\cal N}_{\Psi}(t)$ does not satisfy a key
requirement on the news function, namely its characterization
purely in terms of the geometry of a section ${\cal S}_t$ of the horizon. 
In other words, the flux defined by the square of the news must be
an instantaneous quantity defined by quantities {\em crossing}
the horizon at a given time.
This issue can be addressed
by correcting the integrand in  $\tilde{\cal N}_{\Psi}(t)$
with terms completing $\Psi_0(t')$ to a total time derivative.
From the tidal equation for the evolution of the shear $\sigma^{(h)}_{ab}$
along ${\cal H}$, with leading order given by $\Psi^N_0$,
we propose the more geometric quantity
\bea
\label{e:news_DH}
\frac{dP^{^{({\cal H})}}[\xi]}{dt}(t) =
-\frac{1}{16 \pi}
\oint_{{\cal S}_t} (\xi^i s_i)
\left({\cal N}^{^{({\cal H})}}_{ab}{\cal N}^{^{({\cal H})} ab}\right)
dA  \,, \qquad
{\cal N}^{^{({\cal H})}}_{ab} \equiv - \frac{1}{\sqrt{2}} \sigma^{(h)}_{ab} \ \ .
\eea
The notation $dP^{^{({\cal H})}}[\xi]/dt$ is meant to underline that
this quantity is well defined at ${\cal S}_t$, without 
implying the existence of a well-defined conserved quantity
$P^{^{({\cal H})}}$.

The idea now would be to cross-correlate quantities
$(dP^{^{({\cal H})}}[\xi]/dt)(t)$ at ${\cal H}$ and 
$(dP^{\mathrm{B}}[\xi]/dt)(t)$ at $\scri^+$. This requires however
the determination of the evolution vector $h^a$, which involves
the resolution at each time step of the elliptic Eq. (\ref{e:stability_operator})
for $C$. Although no conceptual issues are involved in this, 
we adopt in a first stage a technically simpler 
strategy where  for $h_{\mathrm{inn}}$, 
rather than  $dP^{^{({\cal H})}}[\xi]/dt$, we use
an {\em effective curvature vector} constructed in terms
of the intrinsic geometry of ${\cal S}_t$.
In order to justify this, we write
\bea
\label{e:evolution_R}
\delta_h {}^2\!R = - \theta^{(h)}\, {}^2\!R 
+ 2 \; {}^2\!D^a{}^2\!D^b  \sigma^{(h)}_{ab}
- {}^2\!\Delta \theta^{(h)} \ ,
\eea
for the evolution of the Ricci scalar ${}^2\!R$ of the induced metric
$q_{ab}$ on ${\cal S}_t$. To fully control its evolution,
we need to track the evolution of $q_{ab}$, $\theta^{(\ell)}$
and $\sigma^{(h)}_{ab}$. In order to get insight into the
involved  quantities, we make explicit the system
for a null horizon (the actual spacelike case
has the same structure, but involving corrections on the
function $C$). Then
\bea
\label{e:evolution_system_horizon}
\delta_\ell {}^2\!R &=& - \theta^{(\ell)}\, {}^2\!R 
+ 2 \; {}^2\!D^a{}^2\!D^b \sigma^{(\ell)}_{ab}
- {}^2\!\Delta \theta^{(\ell)} \nn \\
\delta_\ell q_{ab} &=& 2 \sigma^{(\ell)}_{ab} +  \theta^{(\ell)} q_{ab} \nn \\
\delta_\ell \theta^{(\ell)} &=& -\frac{1}{2} (\theta^{(\ell)})^2
-  \sigma^{(\ell)}_{ab} {\sigma^{(\ell)}}^{ab} - 
8\pi T_{ab}\ell^a\ell^b  \\
\delta_\ell \sigma^{(\ell)}_{ab} &=&  
\sigma^{(\ell)}_{cd} {\sigma^{(\ell)}}^{cd} q_{ab} 
-  {q^c}_a{q^d}_bC_{lcfd}\ell^l\ell^f \nn  \ \ ,
\eea
where ${C^a}_{bcd}$ denotes the Weyl tensor.
 Once initial data are given, the whole system is driven by the {\em external
forces} given by ${q^c}_a{q^d}_bC_{lcfd}\ell^l\ell^f$ and
$T_{ab}\ell^a\ell^b$. Focusing here on the vacuum case
(see \cite{Jaramillo:2011rf} for a more general discussion)
we note that ${q^c}_a{q^d}_bC_{lcfd}\ell^l\ell^f = \Psi_0 \overline{m}_a
\overline{m}_b + \overline{\Psi}_0 m_a m_b$, where
$m^a$ is a complex null vector tangent to ${\cal S}_t$.
On the one hand, the evolution of the 
whole system (\ref{e:evolution_system_horizon})
is determined by the $\Psi_0$ at the horizon, 
which justifies the understanding of ${\cal N}^{^{({\cal H})}}_{ab} $ 
in Eq. (\ref{e:news_DH}) as a kind of news-like function.
On the other hand, we note that the evolution of ${}^2\!R$
is completely driven by the rest of the system, 
without backreacting on it. This last point is crucial, since the evolution of  
${}^2\!R$ then captures in an effective way and in a single
function the evolution of the whole system, in particular
the evolution of the shear (but also other degrees of freedom
if matter is present). This leads to the introduction
of the effective curvature vector \cite{Jaramillo:2011re}
\bea
\label{e:3D_Keff}
\tilde{K}^\mathrm{eff}[\xi](t) \equiv  - \frac{1}{16\pi}
\oint_{{\cal S}_t} (\xi^is_i) |\tilde{\cal N}(t)|^2 dA
\ \ \ \ , \ \ \ \ \tilde{\cal N}(t) \equiv 
\int_{t_c}^t {}^2\!R(t') dt' + \tilde{\cal N}^{t_c} \ \ ,
\eea
as an effective estimator of the evolution of the horizon geometry
(here $\tilde{\cal N}^{t_c}$ is an initial value function; see  \cite{Jaramillo:2011re}
for details on its fixing).
In a first stage of our quantitative analysis, we use 
$\tilde{K}^\mathrm{eff}[\xi]$ as the quantity
$h_{\mathrm{inn}}$ to be cross-correlated with $dP^{\mathrm{B}}[\xi]/dt$.

\begin{figure*}[t]
%\begin{center}
\includegraphics[angle=0, width=5.9cm,clip=true]{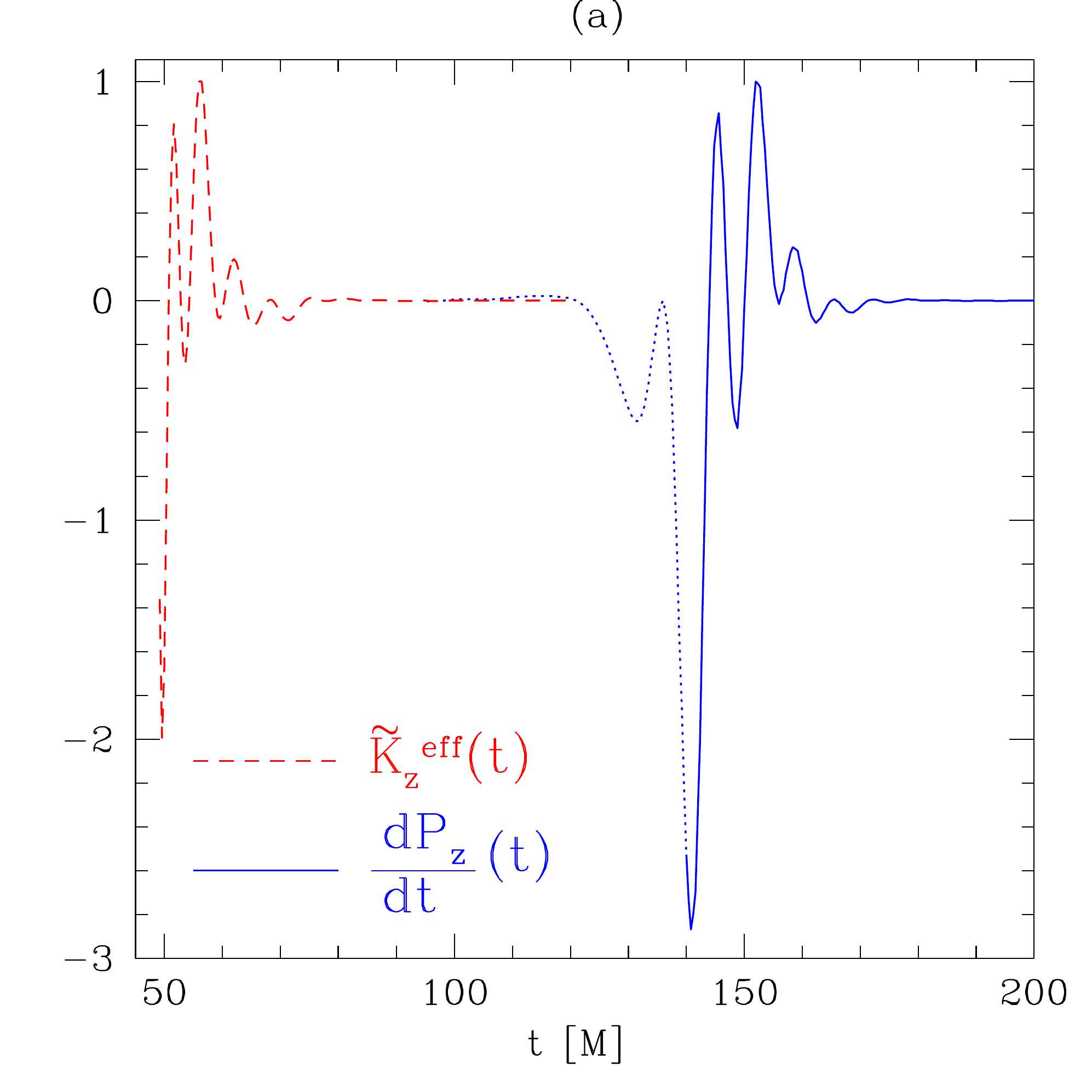}
\includegraphics[angle=0, width=5.9cm,clip=true]{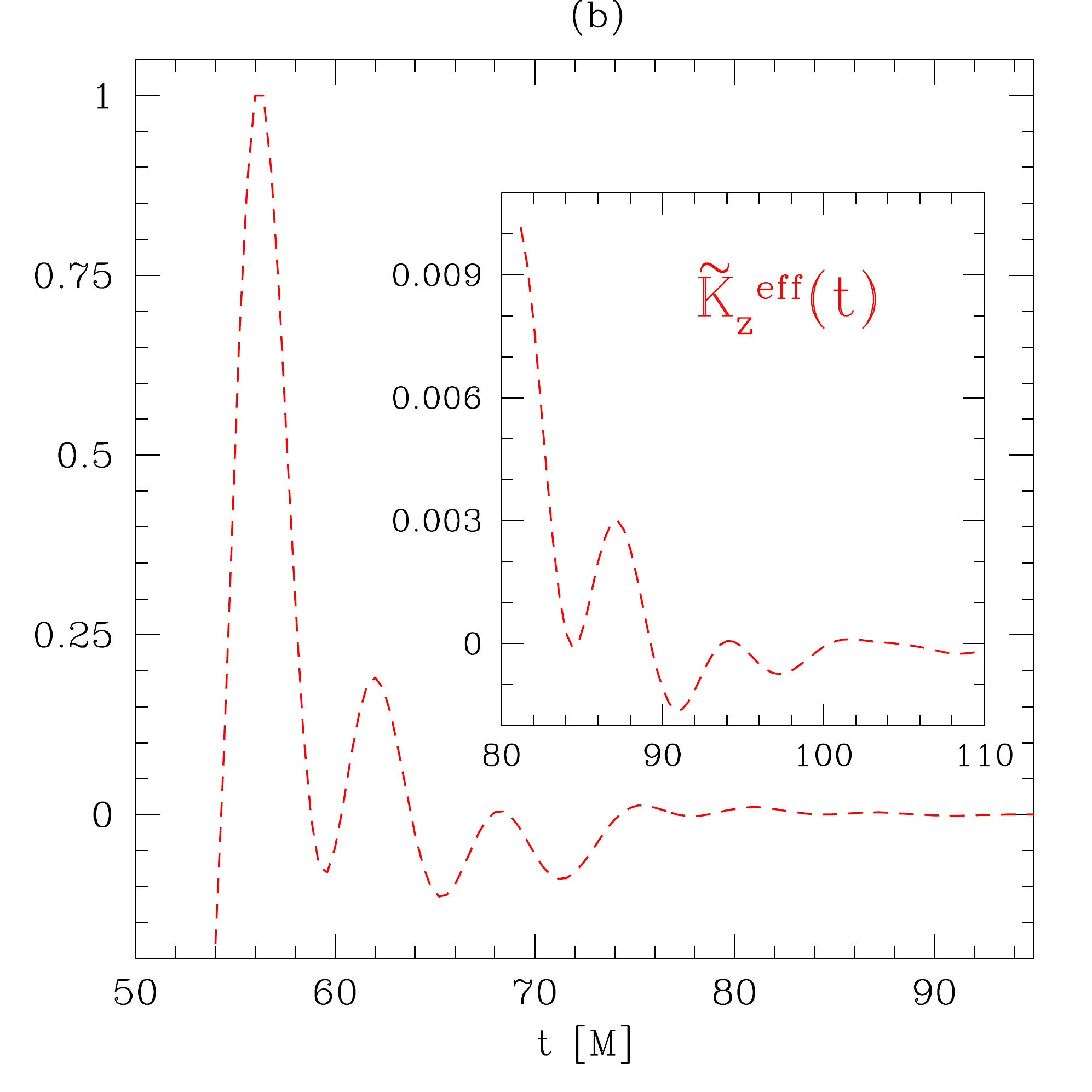}
\includegraphics[angle=0, width=5.9cm,clip=true]{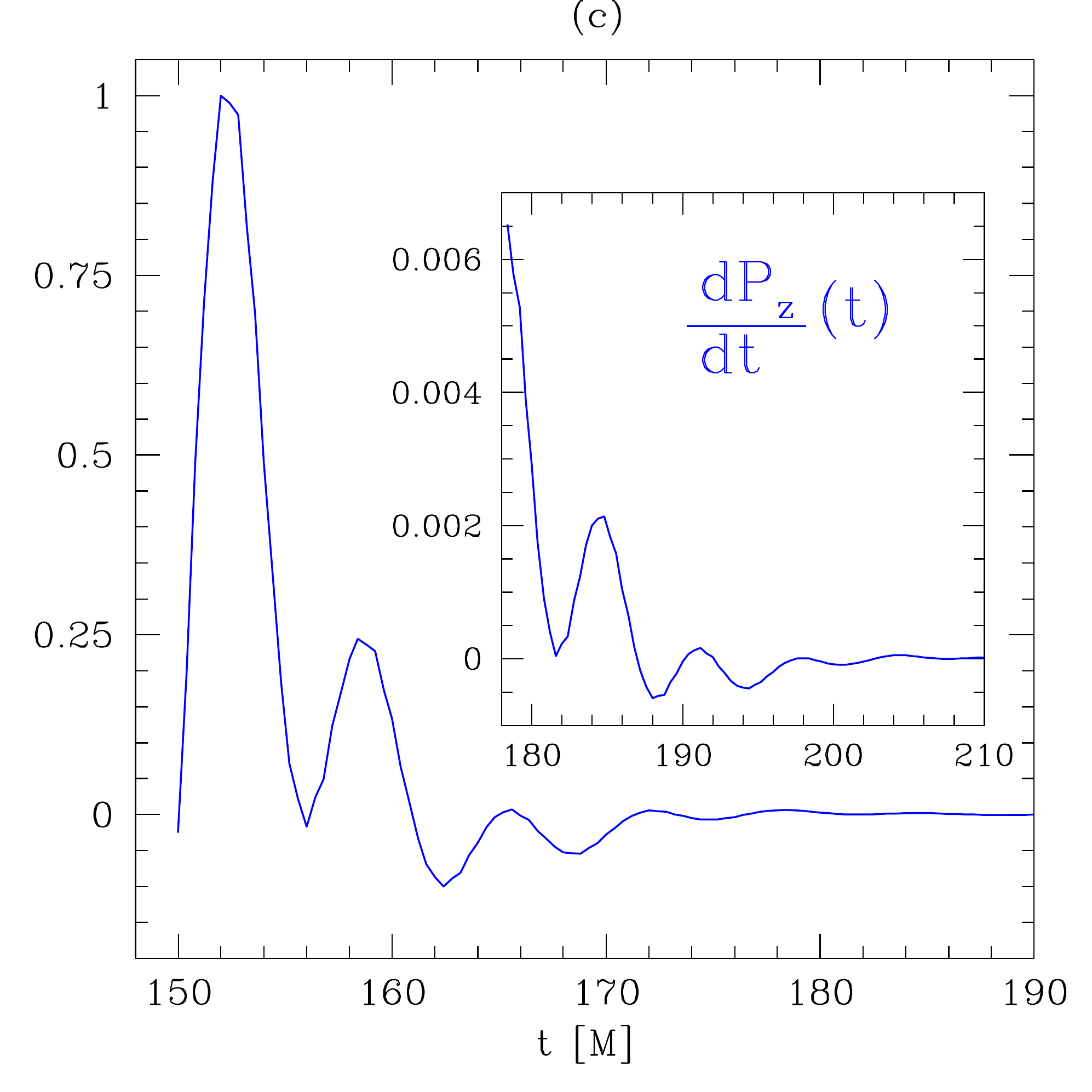}
%\end{center}
%
\caption{Effective curvature $\tilde{K}^{\mathrm{eff}}_z(t)$ at the horizon 
  from the moment of formation of a common horizon
  (red dashed curve) and flux of Bondi linear momentum $(dP^{\mathrm{B}}[\xi]/dt)(t)$
  evaluated at an approximation of  $\scri^+$ (blue dotted solid curves,
  before and after the appearance of a common horizon, respectively).
  Panel (a) shows the good qualitative agreement between both quantities
  after the merger. Panels b) and (c) compare the same quantities for latter times, 
  showing the persistance in time of the good agreement.}  %
\label{fig:flux_time}
\end{figure*}

In order to test these tools, we have considered the head-on collision
of non-spinning black holes with mass ratio  $q=1/2$ and have constructed
numerically the associated dynamical spacetime (cf. details in 
\cite{Jaramillo:2011re}). We extract the timeseries corresponding to 
$\tilde{K}^\mathrm{eff}[\xi](t)$, once the common apparent horizon has formed 
and $(dP^{\mathrm{B}}[\xi]/dt)(t)$ at (an approximation to) null infinity
during the whole evolution.
From a qualitative perspective, both timeseries  show 
a good agreement, from the moment of first appearance of the common
apparent horizon. The qualitative  agreement
is preserved in time (see Fig. \ref{fig:flux_time}).

\begin{figure*}[t]
%\begin{center}
\includegraphics[angle=0, width=5.9cm,clip=true]{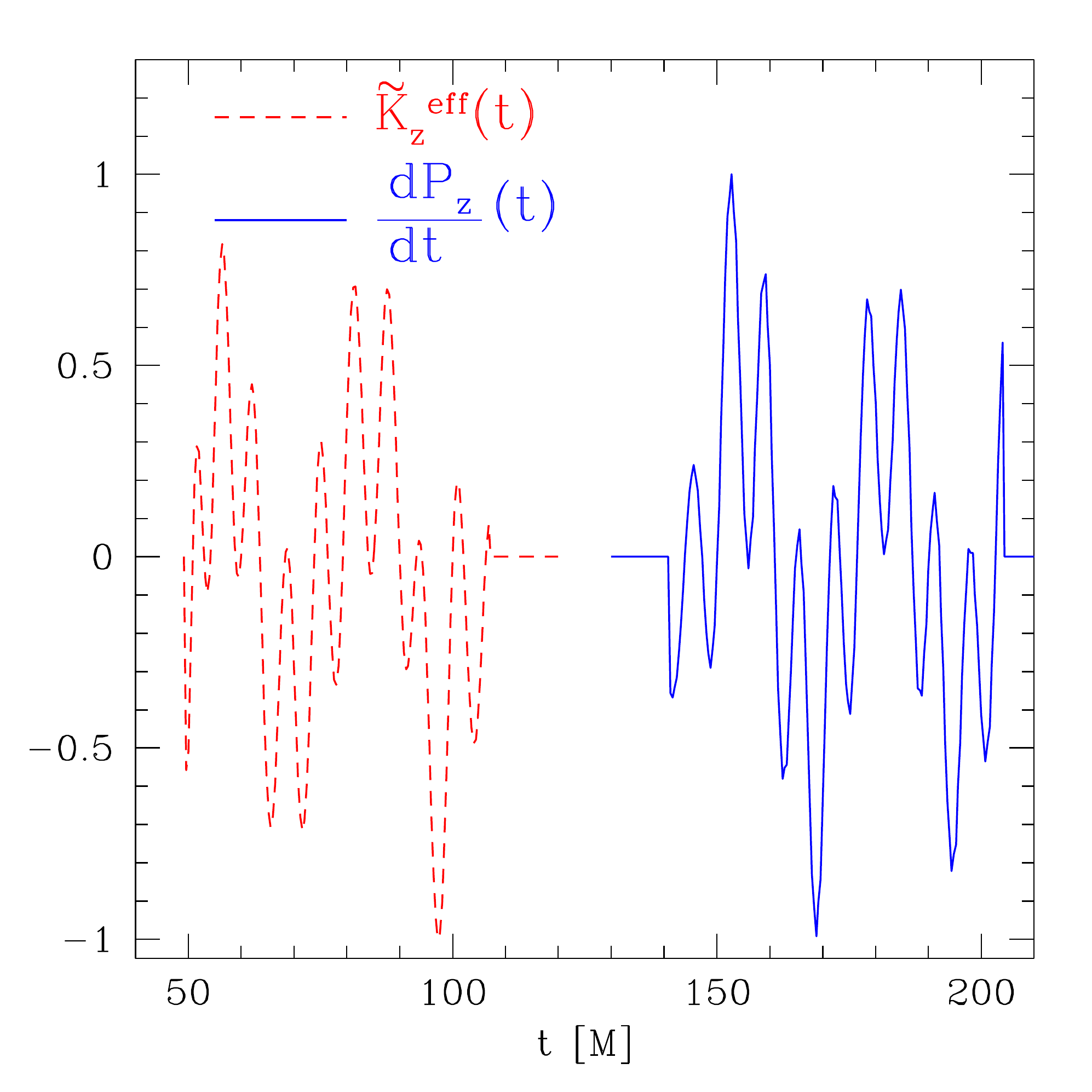}
\includegraphics[angle=0, width=5.9cm,clip=true]{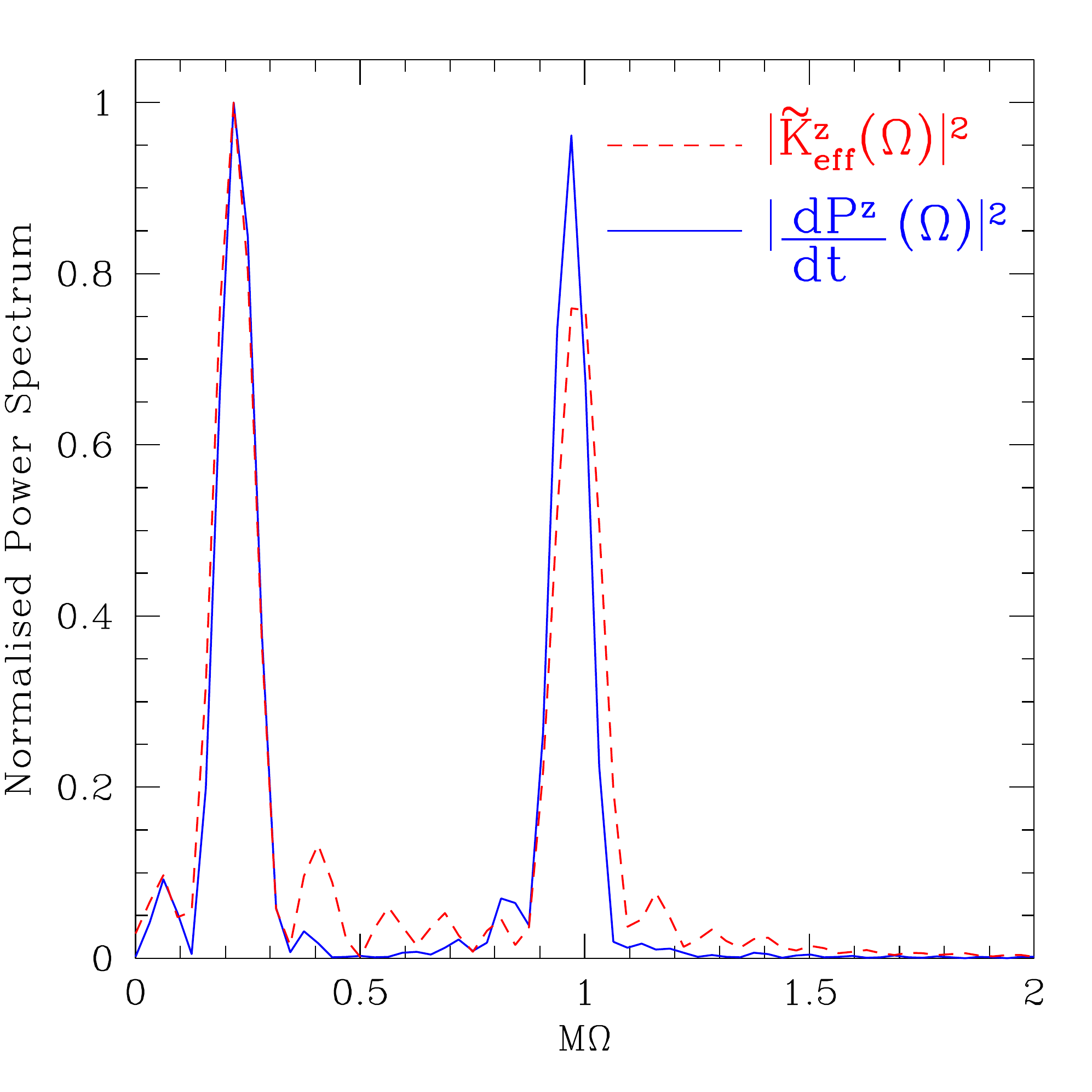}
%\end{center}
%
\caption{
The left panel shows the signals $\tilde{K}^{\mathrm{eff}}_z(t)$ and 
$(dP^{\mathrm{B}}[\xi]/dt)(t)$ shown in Fig. \ref{fig:flux_time},
once the exponential decay has been eliminated. The right
panel presents the power spectrum of the corresponding signals 
showing clearly the presence of two dominating frequencies  
$\Omega^{\mathrm{inn}}_{1,2}$ at ${\cal H}$ and $\Omega^{\mathrm{out}}_{1,2}$  at $\scri^+$.
}
\label{fig:exponential}
\end{figure*}

For a quantitative comparison we use the cross-correlation
of $h_1(t)$ and $h_2(t)$
\bea
\label{e:cross-correlation}
\mathrm{Corr}(h_1,h_2; \tau)&=&\int_{-\infty}^\infty h_1(t+\tau) h_2(t) dt \ \ ,
\eea
which encodes a quantitative comparison
between the two timeseries as a function of the time-shift $\tau$ ({\em lag}) 
between them. 
In particular, the matching between the two signals
can be expressed in terms of the number
\bea
\label{e:matching}
{\cal M}(h_1,h_2)=
\max\limits_{\tau} 
\left(\frac{\mathrm{Corr}(h_1,h_2)(\tau)}
{\left[\mathrm{Corr}(h_1,h_1)(0)\cdot
\mathrm{Corr}(h_2,h_2)(0)\right]^{\frac{1}{2}}}
\right) \ \ .
\eea
This number is confined between $0$ and $1$ (with $1$ indicating
perfect correlation, and $0$ no correlation at all) and provides
the maximum matching between the timeseries.
The calculation of the correlations in our scheme requires  
a careful treatment of what can be referred to as a {\em time stretch issue}, 
in order to deal with the freedom in the choice of spacetime foliation that 
determines the gauge mapping between retarded and advanced times
$u$ and $v$ (cf. \cite{Jaramillo:2011re} for details).
Once this is taken into account, the calculation of the correlation parameter 
in our problem gives typically values ${\cal M}\geq 0.9$, independently of the
width of possible time windows applied to the signals prior to the
calculation of the correlations \cite{Jaramillo:2011re}. 
In order to assess the potential bias in the calculation of the correlations
due to the time decay of the signal, we model the signals by exponentially 
decaying functions, $h_{\mathrm{inn}}(t) = e^{-\kappa_{\mathrm{inn}} t}
h^{\kappa}_{\mathrm{inn}}(t)$ and $ h_{\mathrm{out}}(t)=
e^{-\kappa_{\mathrm{out}} t} h^{\kappa}_{\mathrm{out}}(t)$, and perform the correlation analysis in 
the time series $h^{\kappa}_{\mathrm{inn}}(t)$ and $h^{\kappa}_{\mathrm{out}}(t)$
(cf. left panel in Fig. \ref{fig:exponential}).
Again, the correlation number ${\cal M}\approx 0.9$.
More interestingly, Fourier transforming the signals to get the power spectrum, 
we find that only two frequencies enter the dynamics in this head-on
collision, 
$\Omega^{\mathrm{inn}}_1$ and $\Omega^{\mathrm{inn}}_2$ at ${\cal H}$
and $\Omega^{\mathrm{out}}_1$ and $\Omega^{\mathrm{out}}_2$
at $\scri^+$ (cf. right panel in Fig. \ref{fig:exponential}).
From the very good approximation  \cite{Jaramillo:2011re}
given by $\tilde{K}^{\mathrm{eff}}_z \sim \tilde{\cal N}_2
\tilde{\cal N}_3$ and $(dP_z^{\cal B}/dt) \sim {\cal N}_2 {\cal
  N}_3$, where $\tilde{\cal N}_\ell$ and ${\cal N}_\ell$ are the
($m=0$) spherical harmonic components of $\tilde{\cal N}$ and ${\cal N}$,
respectively.  Remarkably, using a sinusoidal Ansatz for  
$\tilde{\cal N}_\ell$ and ${\cal N}_\ell$, we can reconstruct 
from  $\Omega^{\mathrm{inn}}_{1,2}$ at ${\cal H}$
and $\Omega^{\mathrm{out}}_{1,2}$  at $\scri^+$, the corresponding
$\ell= 2$ and $\ell=3$ modes
\begin{equation}
\label{e:QNM_horizon} 
\Omega^{\tilde{\cal N}}_{\ell=2} = \frac{\Omega^{\mathrm{inn}}_2-
\Omega^{\mathrm{inn}}_1}{2}\,, \quad 
\Omega^{\tilde{\cal N}}_{\ell=3} =
\frac{\Omega^{\mathrm{inn}}_2+\Omega^{\mathrm{inn}}_1}{2}\,, \quad
\Omega^{\cal N}_{\ell=2} = \frac{\Omega^{\mathrm{out}}_2-\Omega^{\mathrm{out}}_1}{2}\,, \quad 
\Omega^{\cal N}_{\ell=3} = \frac{\Omega^{\mathrm{out}}_2+\Omega^{\mathrm{out}}_1}{2}\,,
\end{equation}
leading to an extremely good agreement with 
the real part of the BH quasi-normal mode frequencies 
$\omega^{\rm R}_{\ell=2}$ and $\omega^{\rm R}_{\ell=3}$.
On the other hand, the 
decay inverse time scales $\kappa_{\mathrm{inn}}$ and $\kappa_{\mathrm{out}}$
can be retrieved from the addition of the 
imaginary part of the $\ell=2$ and $\ell=3$ quasi-normal modes:
$\kappa_{\mathrm{decay}} = \omega^{\rm I}_{\ell=2} +  \omega^{\rm I}_{\ell=3}$.
%% \begin{equation}
%% \label{e:QNM_scri_kappa}
%% \kappa_{\mathrm{inn}} = \kappa^{\tilde{\cal N}}_{\ell=2} +
%% \kappa^{\tilde{\cal N}}_{\ell=3} \ \ , \ \ \kappa_{\mathrm{out}} =
%% \kappa^{{\cal N}}_{\ell=2}+\kappa^{{\cal N}}_{\ell=3}. 
%% \end{equation}          
%
This matching of the frequencies at ${\cal H}$ and $\scri^+$
with the quasi-normal modes is shown in Table \ref{t:frequencies_decays}.
On the one hand, this identification of the role played by the quasi-normal 
modes
is consistent with the simple dynamics of the gravitational field in vacuum.
On the other hand, it impacts directly our specific recoil dynamics problem
since characteristic decay and oscillation timescales can be 
constructed from the imaginary and real parts, respectively, of the quasi-normal
modes 
\bea
\tau \equiv \frac{2\pi}{\omega_{\ell=2}^{\textrm{I}} +\omega_{\ell=3}^{\textrm{I}} } \,, 
\qquad
T \equiv \frac{2\pi}{\omega_{\ell=3}^{\textrm{R}} +\omega_{\ell=2}^{\textrm{R}} } \,,
\eea
leading to the slowness parameter introduce in Eq. (\ref{e:slowness_parameter})
\bea
\label{e:P_quasinormal}
P\equiv\frac{T}{\tau}=\frac{\omega_{\ell=2}^{\textrm{I}} +\omega_{\ell=3}^{\textrm{I}} }
 {\omega_{\ell=3}^{\textrm{R}} +\omega_{\ell=2}^{\textrm{R}}} \ .
\eea
%=======================
\begin{table}[]
\caption{The first row shows the comparison between
the $\ell= 2,3$ oscillation frequencies 
at ${\cal H}$ (i.e. $\Omega^{\tilde{\cal N}}_{\ell=2,3}$) and at $\scri^+$ (i.e.
$\Omega^{\cal N}_{\ell=2,3}$) with the real part of  quasi-normal modes 
of a Schwarzschild  BH, $\omega^{\rm R}_{\ell=2,3}$.
The second row shows the comparison between the decay exponents 
$\kappa_{\mathrm{inn}}$ and $\kappa_{\mathrm{out}}$, respectively at  
${\cal H}$ and  $\scri^+$, with the decay coefficient
calculated from the imaginary part of the quasi-normal modes
as $\kappa_{\mathrm{decay}} = \omega^{\rm I}_{\ell=2} +  \omega^{\rm I}_{\ell=3}$.
}
\label{t:frequencies_decays}
%\begin{center}
%\begin{ruledtabular}
\footnotesize{
\begin{tabular}[]{ccc|ccc}
\hline
$M\Omega^{\tilde{\cal N}}_{\ell=2}$ & $M\Omega^{\cal N}_{\ell=2}$ & $M\omega^{\rm R}_{\ell=2}$  & 
$M\Omega^{\tilde{\cal N}}_{\ell=3}$ & $M\Omega^{\cal N}_{\ell=3}$ & $M\omega^{\rm R}_{\ell=3}$ \\
$0.38 \pm 0.04$         & $0.37 \pm 0.04$               & $0.37367$ & 
$0.60 \pm 0.04$         & $0.59 \pm 0.04$               & $0.59944$  \\
&&&&&\\
\hline
&&&&&\\
$M\kappa_{\mathrm{inn}}$   & $M\kappa_{\mathrm{out}}$          & $M\kappa_{\mathrm{decay}}$  
& & & \\
$0.181 \pm 0.006$       & $0.179 \pm 0.005$             & $0.18166$ & & & \\
&&&&&\\
\hline
\end{tabular}
%% \begin{tabular}[]{ccc|ccc}
%% \hline
%% $M\Omega^{\tilde{\cal N}}_{\ell=2}$ & $M\Omega^{\cal N}_{\ell=2}$ & $M\omega^{\rm R}_{\ell=2}$  & 
%% $M\Omega^{\tilde{\cal N}}_{\ell=3}$ & $M\Omega^{\cal N}_{\ell=3}$ & $M\omega^{\rm R}_{\ell=3}$ \\
%% $0.38 \pm 0.04$         & $0.37 \pm 0.04$               & $0.37367$ & 
%% $0.60 \pm 0.04$         & $0.59 \pm 0.04$               & $0.59944$  \\
%% &&&&&\\
%% \hline
%% \end{tabular}
%% \\
%% \begin{tabular}[]{ccc}
%% &&\\
%% $M\kappa_{\mathrm{inn}}$   & $M\kappa_{\mathrm{out}}$          & $M\kappa_{\mathrm{decay}}$  
%% \\
%% $0.181 \pm 0.006$       & $0.179 \pm 0.005$             & $0.18166$ \\
%% &&\\
%% \hline
%% \end{tabular}
}
%\end{ruledtabular}
%\end{center}
\end{table}
%====================================================
This expression has a predictive power to estimate the recoil of the final BH
remnant, from an initial configuration of binary BHs. Indeed, 
using analytic estimations of the
final BH $M$ mass and spin parameter $a$ corresponding to 
given initial configurations \cite{Rezzolla:2008sd},
one can calculate the associated  Kerr quasi-normal modes and construct 
$P$ in (\ref{e:P_quasinormal}). Further insight into
this slowness parameter $P$ is gained from the dynamics of the geometry
of the dynamical horizon ${\cal H}$. As shown in Eq. (\ref{e:evolution_R}), 
the expansion $\theta^{(h)}$ controls the dynamical decay of the
intrinsic geometry. This dissipative role of $\theta^{(h)}$ is further supported
by its interpretation in the membrane paradigm 
\cite{Damou79,Damou82,PriTho86,Price86,GouJar06,Gou05,GouJar06b,GouJar08}
(see also discussion in \cite{Jaramillo:2011rf})
as associated with bulk viscosity terms, whereas $\sigma^{(h)}_{ab}$ is
related to the shear viscosity. The latter is responsible for the 
(shape) oscillations in the intrinsic geometry 
(cf. last equation in (\ref{e:evolution_system_horizon}),
for its relation to propagating gravitational degrees of 
freedom encoded in the Weyl tensor). Given the physical dimensions
$[\theta^{(h)}]=[\sigma^{(h)}]= [\mathrm{Length}]^{-1}$, one could introduce
decay and oscillation inverse timescales 
by averaging $\theta^{(h)}$ and $\sigma^{(h)}$, respectively, over the
horizon section ${\cal S}_t$. In order to make more precise this heuristic 
approach, we consider the evolution equation for $\theta^{(h)}$ \cite{GouJar06b}
\bea
\label{e:evol_theta_h}
\left(\delta_{h} + \theta^{(h)}\right)\theta^{(h)}  &=&
	- \kappa^{(h)} \theta^{(h)} 
	+  \sigma^{(h)}_{ab} {\sigma^{(\tau)}}^{ab} \\
	&&+ \frac{(\theta^{(h)})^2}{2}
	+ {}^2\!D^a ( {}^2\!D_aC- 2C \Omega^{(\ell)}_a ) 
	+ 8\pi T_{ab}\tau^a h^b  
	- \theta^{(k)} \delta_hC \ \ .
\eea
Denoting by $\xi^i_t$ the unit vector in the instantaneous spatial
direction of motion of the BH, decay and oscillation timescales 
can be introduced as
\bea
\label{e:viscosity_timescales2}
\frac{1}{\tau(t)^2}&\equiv& \frac{1}{A}
\oint_{{\cal S}_t}(\xi_t^is_i)\left(\kappa^{(h)}\theta^{(h)}\right) dA \ \nn \\
\frac{1}{T(t)^2}&\equiv& \frac{1}{A}
\oint_{{\cal S}_t}(\xi_t^is_i)
\left(\sigma_{ab}^{(h)}{\sigma^{(\tau)}}^{ab}\right) dA \quad , \quad 
\eea
so that the instantaneous slowness parameter
\bea
\label{e:P_theta_sigma_2}
P(t)=\frac{T(t)}{\tau(t)}=\left(\frac{\oint_{{\cal S}_t}(\xi_t^is_i)
\left(\kappa^{(h)}\theta^{(h)}\right) dA}
{\oint_{{\cal S}_t}(\xi_t^is_i)\left(\sigma_{ab}^{(h)}{\sigma^{(\tau)}}^{ab}\right) 
dA}\right)^{\frac{1}{2}} 
\eea
satisfies $P(t)\approx 1$ near equilibrium, 
when derivative and higher-order terms in (\ref{e:evol_theta_h})  
can be neglected, only surviving the terms in the first line of
the right-hand side, precisely those
used to define $T(t)$ and $\tau(t)$ (note that they lead to the Hartle-Hawking
area evolution equation). Note that $P\approx 1$ is
consistent with the absence of antikick near-equilibrium.

\subsection{Contact with quasi-local BH linear momentum}
Quasi-local notions of linear momentum has
been applied to the study of BH recoil dynamics in 
\cite{KriLouZlo07,Lovelace:2009dg}. From this perspective, the news-like
function ${\cal N}_{ab}^{\cal H}$ in Eq. (\ref{e:news_DH}) can be used
to introduce a heuristic flux of Bondi-like momentum at the horizon
${\cal H}$. Consider the (timelike) unit normal
 $\hat{\tau}^a=\frac{\tau^a}{\sqrt{|\tau^b\tau_b|}}=
\frac{1}{\sqrt{2C}} (\ell^a+Ck^a)=\frac{1}{\sqrt{2C}} (b n^a + N s^a)$ 
to ${\cal H}$ (here, $n^a$ is the timelike unit normal to a 3-slice
$\Sigma_t$ and $s^a$ is the spacelike unit normal to ${\cal S}_t$
in $\Sigma_t$). Consider also
a generic 4-vector $\eta^a$. We can introduce a 
flux of Bondi-like 4-momentum as \cite{Jaramillo:2011rf}
(note the natural use of an advanced time parameter $v$)
\bea
\frac{dP^\tau[\eta]}{dv} \equiv -\frac{1}{8 \pi}
\oint_{{\cal S}_v} (\eta^c \hat{\tau}_c) 
\left({\cal N}^{^{({\cal H})}}_{ab}{\cal N}^{^{({\cal H})} ab}\right)dA 
= -\frac{1}{16 \pi}
\oint_{{\cal S}_v} (\eta^c \hat{\tau}_c) \left({\sigma}^{(h)}_{ab}
{\sigma^{(h)}}^{ab}\right)dA  \ .
\eea
Considering an Eulerian observer $n^a$, the associated flux
of {\em energy} would be
\bea
\frac{dE^{\tau}}{dv}(v) \equiv \frac{dP^\tau[n^a]}{dv}  =
\frac{1}{16 \pi}
\oint_{\cal S} \frac{b}{\sqrt{2C}}\left(\sigma^{(h)}_{ab}{\sigma^{(h)}}^{ab}\right)
dA \ ,
\eea
with $\frac{b}{\sqrt{2C}}=\sqrt{1+N^2/2C}$,
whereas the flux of Bondi-like linear momentum is given by
\bea
\frac{dP^\tau[\xi]}{dv} = -\frac{1}{16 \pi}
\oint_{{\cal S}_v} \frac{N}{\sqrt{2C}}(\xi^i s_i) 
\left(\sigma^{(h)}_{ab}{\sigma^{(h)}}^{ab}\right)dA \ .
\eea
These expressions are closely related to those proposed for DHs \cite{AshKri02,AshKri03}. 
The integration in time $v$ along the horizon ${\cal H}$
provides a heuristic prescription for the linear momentum, 
a sort of Bondi-like counterpart of the heuristic prescription
in Eq. (\ref{e:P_ADM}) based on the ADM momentum. More generally, for
a quantity $Q(v)$ with $F(v)$ flux through ${\cal S}$, we write 
\bea
Q(v)=Q(v_0) +  \int_{v_0}^v F(v')dv' \ \ .
\eea
Note that, in view of the 3+1 description (cf. Fig \ref{f:InnerOuterHor})
such an expression can be split into 
\bea
\label{e:News_t_splitting}
Q(t) =Q_0 
+ \int^{t}_{t_c}\left(F\right)^{\mathrm{int}}(t') dt'
+ \int_{t_c}^t \left(F\right)^{\mathrm{ext}}(t') dt' +
\mathrm{Res}(t)  \nn
\eea
where $\left(F\right)^{\mathrm{int}}$ is the flux through the
internal horizon, $\left(F\right)^{\mathrm{ext}}$ is the flux through 
the external horizon  and 
a residual $\mathrm{Res}(t)= \int^{\infty}_{t}F^{\mathrm{int}}(t') dt'$
must be included
(a more complete expression taking into account changes in the metric type 
of ${\cal H}$ is discussed in \cite{Jaramillo:2011rf}). 
This expression makes explicit the relevance of tracking the internal 
horizon when addressing the integration in time of physical
fluxes across the dynamical horizon ${\cal H}$.

\section{Conclusions and perspectives}

We have outlined some basic elements of a cross-correlation
approach to the analysis near-horizon spacetime dynamics. In particular,
we have identified DHs as hypersurfaces providing inner canonical
screens in a 3+1 Initial Value Problem approach to the spacetime construction,
where appropriate geometric quantities can be defined to probe and monitor
bulk dynamics, namely through correlation 
with quantities at an outer screen.

We have applied this scheme to the study of 
BH recoil dynamics. First, we have 
introduced a heuristic horizon news-like function to build a
quantity $dP^{^{({\cal H})}}[\xi]/dv$
tracking quasi-locally (on ${\cal H}$) the qualitative and to a good extent
also the quantitative features 
of the recoil dynamics at $\scri^+$. In particular, the analysis of its
correlations with the flux of Bondi momentum at $\scri^+$ in numerically
constructed spacetimes corresponding to binary BH head-on collisions 
confirms the simple character of vacuum spacetime dynamics in this setting.
Second, this latter remark has led to the proposal of a prescription for 
a slowness parameter $P$, as constructed from the complex
BH quasi-normal modes. More generally,
inspired by the BH horizon viscous fluid analogy in the membrane paradigm
and further supported by the horizon geometry dynamics, we have proposed
geometric decay and oscillation timescales leading to a more
general characterization of the slowness parameter $P$,
with a well-defined instantaneous meaning.
Third, we have made contact with the heuristic attempts
for estimating the BH 4-momentum, with the proposal of a  
quasi-local Bondi-like expression.
In this context, we have emphasized the relevance of 
the internal 3+1 horizon when considering flux integrations along the BH horizon
${\cal H}$. An open problem in this sense is to assess
the capability of a DH  to {\em dress} the BH singularity,
this involving the understanding of  early and late DH asymptotics.

Despite the encouraging prospects, there are important caveats 
in the sketched cross-correlation approach. 
First, a sound proper formalism is missing.
In particular, an ``inverse scattering'' picture must still be
systematically developed. In addition, the gauge issue briefly
referred to as the ``time stretch issue'' must be addressed in generic
situations. More importantly, it is not clear how to assess  the conditions
under which the comparison/cross-correlation of quantities at
outer and inner screens is actually legitimate.  A possible approach to these
issues (cf. \cite{Jaramillo:2011zw})
would consist in considering the cross-correlation of test-fields
evolving on dynamical spacetimes, without backreacting on them, 
so that the field evolution faithfully tracks specific relevant aspects 
in the geometry of the spacetime dynamics (see also \cite{Bentivegna:2008ei}).
On the one hand, this would remove the ambiguities in the
choice of quantities  $h_{\mathrm {inn}}$ and $h_{\mathrm {out}}$
to be correlated at ${\cal H}_{\mathrm {inn}}$
and ${\cal H}_{\mathrm {out}}$. On the other hand, it would permit to 
extend the scheme to the direct analysis of bulk quantities 
(this would be related to the techniques
in \cite{OweBriChe11,Nichols:2011pu} for tracking spacetime dynamics). 
In particular, the strategy of analyzing the spacetime 
geometry in terms of correlations of appropriate test-fields,
that could be paraphrased as {\em pouring sand on a transparent surface},
can  benefit from the use of tools and concepts 
developed in statistical approaches to field theory in curved
spacetimes, e.g. \cite{Calzetta:1987bw,Hu:1989db,Calzetta:1993qe,Calzetta:1995ea,
Hu:1996gk,Calzetta:1999xh,Hu:2008rga}. 
This attempt leads to the following
declaration of intentions as a perspective for future:
{\em to develop a strategy for spacetime analysis 
aiming at a functional
and coarse-grained description of the spacetime geometry, by importing 
functional tools for the analysis of condensed matter and 
quantum/statistical field theory systems (in curved
backgrounds)}.

%% \begin{figure}
%%   \includegraphics[height=.3\textheight]{golfer}
%%   \caption{Picture to fixed height}
%% \end{figure}

%% %%%%%%%%%%%%%%%%%%%%%%%%%%%%%%%%%%%%%%%%%%%%
%% %% SAMPLE TABLE
%% %%
%% %% Shows the use of \tablehead and \tablenote
%% %% macros
%% %%%%%%%%%%%%%%%%%%%%%%%%%%%%%%%%%%%%%%%%%%%%

%% \begin{table}
%% \begin{tabular}{lrrrr}
%% \hline
%%   & \tablehead{1}{r}{b}{Single\\outlet}
%%   & \tablehead{1}{r}{b}{Small\tablenote{2-9 retail outlets}\\multiple}
%%   & \tablehead{1}{r}{b}{Large\\multiple}
%%   & \tablehead{1}{r}{b}{Total}   \\
%% \hline
%% 1982 & 98 & 129 & 620    & 847\\
%% 1987 & 138 & 176 & 1000  & 1314\\
%% 1991 & 173 & 248 & 1230  & 1651\\
%% 1998\tablenote{predicted} & 200 & 300 & 1500  & 2000\\
%% \hline
%% \end{tabular}
%% \caption{Average turnover per shop: by type
%%   of retail organisation}
%% \label{tab:a}
%% \end{table}

%%%%%%%%%%%%%%%%%%%%%%%%%%%%%%%%%%%%%%%%%%%%%%%%
%% BACKMATTER
%%%%%%%%%%%%%%%%%%%%%%%%%%%%%%%%%%%%%%%%%%%%%%%%

\begin{theacknowledgments}
We would like to thank the organizers of the Spanish Relativity Meeting, ERE2011.
\end{theacknowledgments}



\bibliographystyle{aipproc}   % if natbib is available

\bibliography{ERE2011}

\IfFileExists{\jobname.bbl}{}
 {\typeout{}
  \typeout{******************************************}
  \typeout{** Please run "bibtex \jobname" to optain}
  \typeout{** the bibliography and then re-run LaTeX}
  \typeout{** twice to fix the references!}
  \typeout{******************************************}
  \typeout{}
 }

\end{document}